\documentclass[preprint,12pt]{elsart}
\usepackage{graphicx,amssymb,amsmath,times}
\usepackage{setspace}
\journal{Journal of High Energy Astrophysics}

\begin{document}
\begin{frontmatter}
\title{Gamma-ray and Optical properties of the flat spectrum radio quasar 3C 279 flare in June 2015}
\author[1,2]{K K Singh\corauthref{cor}},
\corauth[cor]{Corresponding author.}
\ead{kksastro@barc.gov.in}
	\author[1]{P J Meintjes}, \author[1]{B Bisschoff}, \author[1]{F A Ramamonjisoa}, \author[1]{B van Soelen }
\address[1]{Physics Department, University of the Free State, Bloemfontein- 9300, South Africa} 
\address[2]{Astrophysical Sciences Division, Bhabha Atomic Research Centre, Mumbai- 400 085, India}

\begin{abstract}
The flat spectrum radio quasar 3C 279 was observed in an extremely high activity state on June 16, 2015 (MJD 57189). 
In this paper, we investigate the properties of this flaring episode in the high energy $\gamma$-ray and optical 
bands using data from the \emph{Fermi}-LAT, SMARTS, and SPOL observations during the period June 1-30, 2015 (MJD 57174-57203). 
The highest emission state in the $\gamma$-ray band detected by the \emph{Fermi}-LAT exhibits a peak flux of 
$\sim$ 2$\times$10$^{-7}$ erg~cm$^{-2}$~s$^{-1}$ which is more than 25 times the flux level measured in the low activity 
state of the source. The temporal analysis of the daily \emph{Fermi}-LAT light curve suggests that the giant flaring episode 
has characteristic rise and decay times less than one day. The optical daily light curves in B, V, R, and J bands 
also indicate the flaring activity from 3C 279 with flux levels peaking for two days on June 16-17, 2015 
(MJD 57189-57190). The discrete correlation function analysis indicates a time lag of 1 day or longer between the 
$\gamma$-ray and optical peaks during the flaring episode. The $\gamma$-ray emission is also observed to show 
a harder-when-brighter behaviour whereas optical emission exhibits an opposite behaviour. The $\gamma$-ray emission 
region during the flare is observed to be very compact and is located close to the base of the jet. The degree of 
linear polarization in the wavelength range 500-700 nm measured using SPOL during this period is also highly variable 
with a peak of $\sim$ 30$\%$ one day after the $\gamma$-ray flare. Near simultaneous $\gamma$-ray flux points show a 
linear anti-correlation with the degree of polarization during the period of $\gamma$-ray flare. The significant drop in 
the degree of linear polarization suggests a sudden increase in the tangled magnetic field strength in the emission region. 
\end{abstract}
\begin{keyword}
(Galaxies:) Flat Spectrum Radio Quasars: individual: 3C 279: data analysis-Gamma-rays: general:optical-SMARTS
\end{keyword}

\end{frontmatter}

\section{Introduction}  
Flat spectrum radio quasars (FSRQs) are subclass of blazars (radio-loud active galactic nuclei having relativistic 
jets pointing in the direction of the observer at the Earth) with prominent and broad emission lines in the optical 
spectrum [1,2]. These objects are observed to be more luminous than another class of blazars namely BL Lacertae objects 
(BL Lacs). The immense luminosity of blazars (including FSRQs and BL Lacs) is attributed to the release of  gravitational 
energy through the accretion onto the supermassive black hole (SMBH) and to the relativistic jets produced by the rotation 
of the spinning black hole at the centre [3,4]. However, the launching mechanism of the collimated relativistic jets from 
the central region of the host galaxy is not fully understood. The relativistic jets transport energy and momentum to very 
large scales and are assumed to be the most promising sites for non-thermal multi-wavelength emission from blazars. 
Optical spectroscopy of a large sample of FSRQs reveals that the optical continuum from these objects is dominated by 
non-thermal emission associated with the relativistic jet [5]. The degree of non-thermal dominance is correlated with 
the high energy (HE; E $>$ 0.1 GeV) $\gamma$-ray spectral index [5,6]. However, radio emissions do not show any strong 
correlation with the optical spectral properties of the FSRQs [5]. A high degree of linear polarization in the optical 
band measured from the FSRQs indicates the presence of a well ordered magnetic field in the jet [7]. The observation 
of any coincidence between changes in the $\gamma$-ray activity with the optical emissions and polarization provides 
evidence for the site of emission regions in the jet. The coincidence of a $\gamma$-ray flare with a significant 
change in the angle of optical polarization indicates co-spatiality of $\gamma$-ray and optical emission regions in 
the jet [8]. 
\par
Continuum broad-band radiation from radio to very high energy (VHE; E $>$ 100 GeV) measured from the jet can be described 
by a spectral energy distribution (SED) with two characteristic humps peaking at low and high frequencies. The low energy 
spectral component in the SED of FSRQs peaks at IR/optical frequencies and is attributed to the synchotron radiation of 
relativistic eletrons in the jet magnetic field [9,10]. The cut-off frequency of synchrotron radiation in 
the SED characterizes the maximum energy to which the electrons can be accelerated in the jet [11]. The second hump in 
the SED of FSRQs has a peak frequency ranging between  hard X-ray and HE $\gamma$-rays and its origin is 
conventionally explained either by the external inverse Compton (EIC) scattering in the leptonic scenario or by the 
hadronic processes involving proton synchrotron and electromagnetic cascade emission initiated by the $p\gamma$ interactions [12]. 
The target photons for the EIC scattering by the relativistic electrons in the emission region of the jet are thermal photons 
from a bright accretion disk, IR photons from a dusty torus surrounding the SMBH, scattered radiation from broad-line regions, 
photons from the sheath region of the jet or photons from the cosmic microwave background radiation [13]. However, the exact 
origin of the external radiation field is still being debated in blazar research. On the other hand, protons can be accelerated to 
relativistic energies by different physical processes (e.g. stochastic acceleration or diffusive shock acceleration) 
in the jet and produce HE $\gamma$-rays via hadronic processes [14]. The relativistic protons interact either with the synchrotron 
photons produced by the co-accelerated electrons in the emission region or with the photons coming from the regions external to the 
jet and initiate electromagnetic cascade [15]. This process predicts comparable $\gamma$-ray and neutrino fluxes from the FSRQs [16]. 
The emission of $\gamma$-rays through proton synchrotron process requires a very strong magnetic field in the jet and does not 
necessarily predict neutrino production [17,18]. Recent evidence for the astrophysical neutrino events from TXS 0506+056 
during the flaring activity provides a smoking gun signature of the hadronic processes in the blazar jet [19,20]. 
However, the broad-band SED of the source with a synchrotron peak frequency less than 10$^{14}$ Hz and a dip in the X-ray band 
challenges the hadronic process based on the electromagnetic cascading because it is not able to explain the observed $\gamma$-rays [21]. 
Whereas, the hadronic processes including proton synchrotron and photo-meson production can explain the $\gamma$-ray component 
in the SED but the predicted neutrino flux is much below the observed flux level [21]. Therefore, the exact $\gamma$-ray emission 
mechanism in FSRQs and blazars in general remains an open question in astrophysics research. 
\par
The observed HE $\gamma$-ray emissions from FSRQs are highly variable at various timescales with extreme properties during the 
flaring episodes when the flux level attains a peak value which is several times higher than the flux measured during low activity state. 
The bolometric $\gamma$-ray luminosity above 0.1 GeV is observed up to $\sim$ 10$^{48}$ erg~s$^{-1}$ and the jet luminosity is 
found up to $\sim$ 10 times the disk luminosity [22]. The broad-band SED is strongly Compton dominated with the observed 
$\gamma$-ray luminosity being much higher than the synchrotron luminosity and the $\gamma$-ray light curves are weakly correlated with 
low energy emissions in the  optical and soft X-ray bands. A large population of FSRQs shows such observational features but the 
physical processes involved are not clearly understood. In the present study, we explore the properties of a giant $\gamma$-ray outburst 
observed from the FSRQ 3C 279 in June 2015. The paper is structured as follows: Section 2 summarizes the important observational features 
of the flaring episodes from 3C 279. In Section 3, we describe the data set from $\gamma$-ray and optical observations used in this work. 
The results are discussed in Section 4. Some of the important properties of the emission region in the jet of 3C 279 are derived in Section 5. 
Finally, we conclude the important findings of the study in Section 6.

\section{3C 279}
3C 279 is an FSRQ hosted by a galaxy at redshift $z$= 0.5362 harbouring a central SMBH of mass $\sim$ 8$\times$10$^8$ M$_\odot$ [23,24]. 
Radio observations of this FSRQ suggest that the relativistic jet is extended up to kiloparsec (kpc) scales with a viewing angle 
$\sim$ 2$^\circ$ [25,26]. 3C 279 is the first object in the FSRQ subclass which has shown a strong detection of variable HE $\gamma$-ray 
emission in the energy range 30 MeV - 5 GeV in June 1991 with an integral flux of (2.8$\pm$0.4)$\times$10$^{-6}$ ph~cm$^{-2}$~s$^{-1}$ 
above 0.1 GeV [27]. In February 2006, a giant VHE $\gamma$-ray flare from 3C 279 was detected by a ground-based Cherenkov telescope 
above an energy threshold of 50 GeV and the integral flux above 100 GeV was found to be 
(5.15$\pm$0.82)$\times$10$^{-10}$ ph~cm$^{-2}$~s$^{-1}$ [28]. During this flare, the optical emission in the R band was also observed to 
increase by a factor of 2 above the low activity state without any short timescale variability [28]. A sharp HE $\gamma$-ray flare was 
detected with a doubling timescale of less than one day in February 2009 [8]. During this flaring episode, the degree of optical linear 
polarization dropped significantly whereas the fluxes in different optical bands were relatively high and variable [8]. Near simultaneous 
X-ray observations indicated a low or steady emission state of 3C 279. But an X-ray flare was observed after about 60 days for a period 
of $\sim$ 20 days similar to the duration of HE $\gamma$-ray flaring episode. 
Multiple distinct HE $\gamma$-ray flares with a peak flux level of 10$^{-5}$ ph~cm$^{-2}$~s$^{-1}$ above 100 MeV and flux doubling 
timescale of 2 hours were observed from 3C 279 between December 2013 and April 2014 [29,30]. 
Another giant flaring activity in HE $\gamma$-ray band with the historically highest peak flux of 
3.6$\times$10$^{-5}$ ph~cm$^{-2}$~s$^{-1}$ above 100 MeV from this source was detected in June 2015 [31,32]. 
The $\gamma$-ray flux doubling time estimates during this flaring episode are less than 5 minutes. 
This flaring event was also detected by the High Energy Stereoscopic System (H.E.S.S.) above an energy threshold of 66 GeV [33]. 
During the night of June 16, 2015 observations, CT5 telescope in the H.E.S.S. array detected VHE $\gamma$-ray photons from 3C 279 with  
a statistical significance of 8.7$\sigma$ in $\sim$ 2.2 hours of observation. The average VHE flux above 200 GeV was estimated to be 
(7.6$\pm$0.7)$\times$10$^{-12}$ ph~cm$^{-2}$~s$^{-1}$ [33]. Results from the multifrequency follow-up observations including X-ray and 
optical data on this giant $\gamma$-ray flare from 3C 279 are discussed in [34]. Apart from the giant outbursts frequently detected from 
the FSRQ 3C 279 in HE $\gamma$-rays, several multi-wavelength studies of the source during low activity state over three decades are 
also reported [35,36,37,38,39,40]. However, the physical processes involved in the flaring and the low activity states as well as 
the behaviour of the broad-band emissions from 3C 279 remain poorly understood.

\section{Data Set}
In this study, we focus on the strongest HE $\gamma$-ray outburst observed from the FSRQ 3C 279 in June 2015 [31,32,33,34,41]. We have 
used data from the near simultaneous HE $\gamma$-ray and optical observations of the source during the period June 1-30, 2015 
(MJD 57174-57203). The details of the data from different instruments and their reduction procedure followed in the present work are 
described below:    

\subsection{\emph{Fermi}-LAT}
The HE $\gamma$-ray data in the energy range 0.1-500 GeV have been used from the \emph{Fermi}-Large Area Telescope (LAT) 
observations of 3C 279 for the period June 1-30, 2015 (MJD 57174-57203). The LAT is a pair-conversion $\gamma$-ray detector on 
board  the \emph{Fermi} satellite and provides observations of the astrophysical sources after every 3 hours in survey mode [42]. 
We have analyzed the publicly available Pass8 (P8R3) data \footnote{https://fermi.gsfc.nasa.gov/cgi-bin/ssc/LAT/LATDataQuery.cgi}  
following the standard procedure using \emph{Fermi}tools version 1.0.1 (Fermi 1.0.1) 
software\footnote{https://fermi.gsfc.nasa.gov/ssc/data/analysis/documentation/Cicerone}. 
We have extracted the photon-like events belonging to the SOURCE class (evclass=128 \& evtype=3) in the 
energy range 0.1-500 GeV. An unbinned likelihood analysis is performed for a region of interest (ROI) with 10$^\circ$ radius centred at 
the position of the FSRQ 3C 279 (R.A.=195.048$^\circ$, Decl.= -5.788$^\circ$). All the point sources within 15$^\circ$ from the position 
of 3C 279 are included from the fourth \emph{Fermi}-LAT source (4FGL) catalogue [43]. The spectrum of the source under study is defined 
as a log-parabola function 
\begin{equation}\label{eqn:lp}
	\frac{dN}{dE}~=~K~\left(\frac{E}{E_0}\right)^{-\alpha - \beta~\ln (E/E_0)}
\end{equation}	
in the 4FGL catalogue, where $K$ (normalization), $\alpha$ (spectral slope at pivot energy $E_0$) and $\beta$ (curvature index) are 
the parameters in the model. The daily light curves for 3C 279 are generated using a maximum likelihood algorithm implemented in 
\emph{gtlike}. As on short timescales the statistics may not be sufficient to detect the curvature, the spectrum of 
the source 3C 279 has been fitted with a power law function ($\beta =0$) to obtain the daily light curve, whereas the spectra of 
the remaining point sources are kept the same as defined in the 4FGL catalogue. The spectral parameters of all the sources within 
the ROI vary in the energy range 100 MeV to 500 GeV. For the sources outside the ROI, the spectral parameters are freezed to 
the 4FGL values during fitting. The Galactic diffuse and extragalactic isotropic emissions are modelled using the background 
model files  $gll\_iem\_v06.fits$ and $iso\_P8R3\_SOURCE\_V2.txt$ respectively.
In order to quantify the significance of the HE $\gamma$-ray photons detected from the source, a Test Statistic (TS) is estimated from the 
ratio of the maximum value of the likelihood function over the ROI including and without the source in the model respectively [44]. 
We have considered a detection significance threshold of TS $>$ 25 (corresponding to the statistical significance above 5$\sigma$) in 
the present study. The daily light curve of 3C 279 obtained in the present work is compatible with the one already reported  
in [31] over the energy range 100 MeV to 300 GeV.

\subsection{SMARTS}
The Small and Moderate Aperture Research Telescope System (SMARTS) at  Cerro Tololo Interamerican 
Observatory (CTIO)\footnote{http://www.astro.yale.edu/smarts/glast/home.php} provides simultaneous optical/IR observations 
of the \emph{Fermi} blazars in a wide wavelength range from 400 nm to 2200 nm under the Multiwavelength Observing-Support 
Programs\footnote{https://fermi.gsfc.nasa.gov/ssc/observations/multi/programs.html}. The observation is carried out with 
the SMARTS 1.3m and ANDICAM instrument [45]. ANDICAM is a dual-channel imager with a dichroic comprising of an optical CCD 
and an IR imager and provides simultaneous observations in B, V, R, and J bands with greatest coverage. Comparison stars 
are calibrated in the field to determine the magnitudes of the target source in different bands. More details of the data 
analysis can be found in [46]. We have used the publicly available online data of 3C 279 for the period June 1-30, 2015 
(MJD 57174-57203). The observed magnitudes in different bands are dereddened to account for the absorption along the line 
of sight (Galactic extinction) before converting to appropriate flux units for comparison with other multi-wavelength observations. 
We have used the following relation for converting the observed magnitudes ($m_0$) into energy flux densities [47,48] 
\begin{equation}
	F_\nu~(\rm{erg~cm^{-2}~s^{-1}~Hz^{-1}})~=~F_0~10^{-0.4(m_0 + A_{\lambda})}
\end{equation}	
where $A_\lambda$ is the Galactic extinction in a given band and $F_0$ is the corresponding zero magnitude flux density. 
The values of these parameters which are used to estimate the energy flux in the daily light curves for four bands 
(B, V, R, and J) are given in Table \ref{tab:opt-conv}.
\begin{table}
\caption{Galactic extinction and zero magnitude flux values in different optical/IR bands for the conversion of observed magnitudes 
	from SMARTS observations into energy fluxes [49,50].}
\begin{center}
\begin{tabular}{cccc}
\\
\hline
Band 		&Effective frequency ($\nu$)	&$A_\lambda$ 	&$F_0$ (10$^{-20}$ erg~cm$^{-2}$~s$^{-1}$~Hz$^{-1}$)\\
\hline
B		&6.73$\times$10$^{14}$Hz	&0.123		&4.063\\			
V		&5.44$\times$10$^{14}$Hz	&0.093		&3.636\\
R		&4.55$\times$10$^{14}$Hz	&0.075		&3.064\\
J 		&2.45$\times$10$^{14}$Hz 	&0.027		&1.589\\
\hline
\end{tabular}
\end{center}
\label{tab:opt-conv}
\end{table}
\subsection{SPOL}
The Spectro-Polarimeter (SPOL) at Steward Observatory of the University of Arizona \footnote{http://james.as.arizona.edu/psmith/Fermi/} 
contributes optical linear polarization data in the wavelength range from 400 nm to 700 nm for a large sample of LAT monitored blazars 
under \emph{Fermi} Multiwavelength Observing-Support Programs [51]. The polarimeter in the SPOL instrument has a dual beam design 
which incorporates a full-aperture Wollaston prism and rotating achromatic waveplates. It can be mounted directly on the  2.3m 
Bok Telescope and 1.54 m Kuiper Telescope as a standalone instrument. More details about the design of the SPOL can be obtained 
in [52]. A $\lambda$/2 waveplate is inserted into the telescope beam to measure the linear polarization at a level less than 0.05$\%$. 
The degree of optical linear polarization ($P$) and angle of polarization ($\phi$) yielded from the SPOL observations are given by 
\begin{equation}
	P(\%)~=~100 \times \sqrt{q^2 + u^2}~~~~;~~~~~~ \phi(^\circ)~=~\frac{1}{2}~~\arctan\left(\frac{u}{q}\right)
\end{equation}
where $q$ and $u$ are the mean values of the normalized Stokes parameters for the imaging data. The value of $P$ is also corrected for the 
statistical bias associated with it being a definite quantity and $P > 0$ [53]. However, no interstellar polarization correction has  
been made to the values of $P$. Also, $\phi$ is measured in the astronomical sense that its value increases going east of north on the sky 
and $\phi$ =0$^\circ$ or 180$^\circ$ indicates that the polarization vector is aligned in the direction of north-south. 
The measurement of $\phi$ has a 180$^\circ$ ambiguity with $\phi$ =90$^\circ$ or 270$^\circ$ being the same indicating an east-west 
orientation of the linear polarization vector. We have used the publicly available daily values of $P$ and $\phi$ during the period 
June 1-30, 2015 (MJD 57174-57203) in the present study. We have also used the archival SPOL-photometric data in V and R bands 
available for the periods missing in the SMARTS observations.

\begin{figure}
\begin{center}
\includegraphics[width=1.0\textwidth]{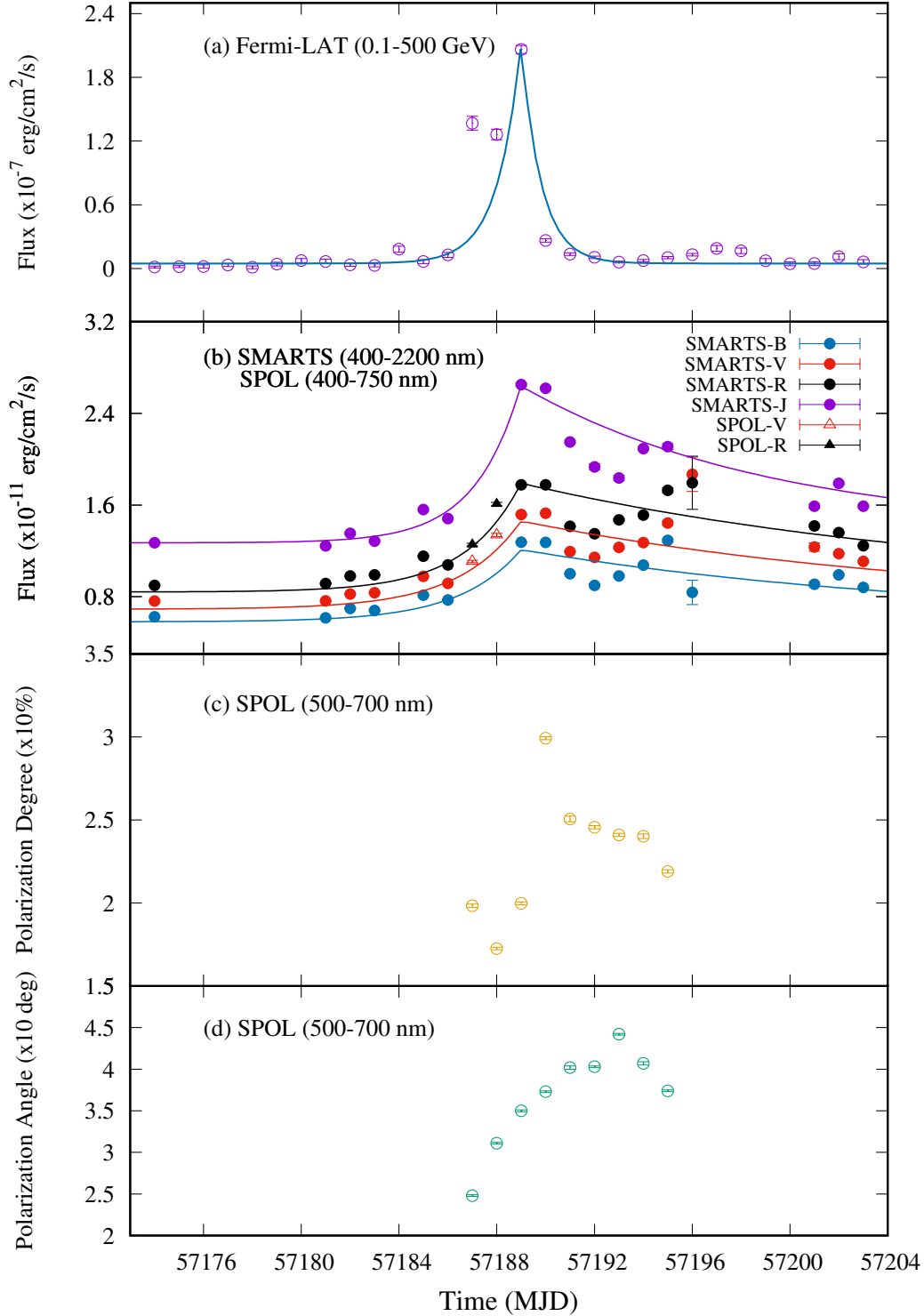}
\caption{HE $\gamma$-ray and optical light curves of the FSRQ 3C 279 during June 1-30, 2015 (MJD 57174-57203). The flux points shown  
	in the panels (a) and (b) have been averaged over one day. Curves in the panels (a) and (b) are the temporal profile of the 
	light curves obtained from the fitting using Equation \ref{eqn:temporal}. In the panels (c) and (d), the degree of linear 
	polarization and the corresponding angle of polarization for daily observations are reported.}
\label{fig:Fig1}
\end{center}
\end{figure}
\section{Results and Discussion}
We have obtained the HE $\gamma$-ray and optical daily light curves from the \emph{Fermi}-LAT and SMARTS/SPOL observations of the 
FSRQ 3C 279 respectively for the period June 1-30, 2015 (MJD 57174-57203). Apart from the flux measurements, we also use near 
simultaneous optical polarization measurements which are available from the SPOL observations. Important findings from these 
data sets during the strongest flaring episode of 3C 279 are discussed in the following sub-sections.

\subsection{Light curves}
The daily light curves obtained from the near simultaneous HE $\gamma$-ray and optical observations of the FSRQ 3C 279 are shown in 
Figure \ref{fig:Fig1} (a) and (b) respectively. We see from Figure \ref{fig:Fig1} (a) that the $\gamma$-ray emission from the 
source is consistent with a low activity state during the period June 1-13, 2015 (MJD 57174-57186) and suddenly starts increasing 
to attain a peak value on June 16, 2015 (MJD 57189), and immediately decreases to remain in the low activity state during 
June 18-30, 2015 (MJD 57191-57203). The HE $\gamma$-ray flux increases significantly on June 14, 2015 (MJD 57187) with 
respect to the low activity state and remains in the high activity state for two days (MJD 57187-57188) before reaching 
the highest flux level on June 16, 2015 (MJD 57189). The low activity states before and after the peak are described by 
a constant flux levels of (3.40$\pm$0.92)$\times$10$^{-9}$ erg~cm$^{-2}$~s$^{-1}$ and 
(1.01$\pm$0.14)$\times$10$^{-8}$ erg~cm$^{-2}$~s$^{-1}$ respectively. Therefore, the low activity state of the source during 
June 2015 has an average flux level of (6.97$\pm$1.07)$\times$10$^{-9}$ erg~cm$^{-2}$~s$^{-1}$ above 100 MeV. The peak flux during 
the highest activity state on June 16, 2015 (MJD 57189) is (2.06$\pm$0.04)$\times$10$^{-7}$ erg~cm$^{-2}$~s$^{-1}$ which is more 
than 25 times the average flux level in the low activity state and about 1.5 times the high flux level measured during the 
first two days before the peak. The highest activity state in the HE $\gamma$-ray band detected by the \emph{Fermi}-LAT is also 
accompanied by near simultaneous increase in the optical emission from 3C 279 as shown in Figure \ref{fig:Fig1} (b). 
From the optical light curves in four bands (B, V, R, and J) obtained from the SMARTS and SPOL observations, 
it is evident that the increase in the flux level during the highest activity with respect to the low emission state is not as 
high as in the HE $\gamma$-ray emission of the source. The optical emission remains in the relatively high activity state 
for one more day after the sudden decay of the HE $\gamma$-ray outburst. However, the optical emissions in B, V, and R bands 
again start to increase after the decay of the flare and attain a second peak on June 23, 2015 (MJD 57196). The observed 
flux levels in two optical bands (B and R) during the second peak are similar to those corresponding to the flare on 
June 16-17, 2015 (MJD 57189-57190). Whereas, the flux level in V band on June 23, 2015 (MJD 57196) is higher than the 
flux points measured during June 16-17, 2015 (MJD 57189-57190). All three data points corresponding to the second peak exhibit relatively 
large error bars and are also not consistent with the overall trend in the optical light curves in Figure \ref{fig:Fig1}(b). 
No such activity is observed in the HE $\gamma$-ray band. The flux doubling timescale ($t_d$) during the flaring activity is 
estimated using the relation [54]
\begin{equation}\label{eqn:doub-time}
	t_d = \frac{t_2 - t_1}{\log_2 (F_2/F_1)}		
\end{equation}	
where $F_1$ and $F_2$ are the flux values at consecutive times $t_1$ and $t_2$ respectively in a given light curve. 
A one day binned near simultaneous HE $\gamma$-ray and optical light curves (Figure \ref{fig:Fig1} (a) \& (b)) reveal 
a flux doubling timescale of $\sim$ 18 hours and $\sim$ 4 days respectively during the flaring episode. 
This indicates that the variation in the $\gamma$-ray emission in the energy range 0.1-500 GeV of 3C 279 during the 
flare is stronger than the emissions in different optical bands. However, it is important to note that the above estimates 
for the flux doubling timescale are affected by the one day binning of the light curves. A finer binning of the 
HE $\gamma$-ray light curve gives a much shorter flux doubling timescale of $\sim$ 2 hours during the extreme flaring activity 
on June 16, 2015 (MJD 57189) of the FSRQ 3C 279 [31]. Ackermann et al. (2016) have estimated a much shorter $\gamma$-ray flux 
doubling timescale of less than 5 minutes during this outburst using the \emph{Fermi}-LAT light curves which are binned 
over 3-5 minutes [32].
\par
To characterize the temporal profile of the flaring activity observed in the HE $\gamma$-ray and optical bands, we have fitted 
the flux points in the light curves which are reported in Figure \ref{fig:Fig1} (a) and (b) using the function [55]
\begin{equation}\label{eqn:temporal} 
	F(t) = F_0 + F \left[e^{(t-t_p)/\tau_r} H(t_p-t) + e^{-(t-t_p)/\tau_d } H(t-t_p) \right]
\end{equation}
where $\tau_r$ and $\tau_d$ are the rise and decay timescales respectively and $t_p$ is the time of highest flux. $F_0$ 
and $F$ are the parameters  for determining the flux levels during low and high activity states respectively and $H$ 
is the \emph{Heaviside function}. The HE $\gamma$--ray and optical light curves are fitted using 
$\tau_r$, $\tau_d$, $F_0$ and $F$ as parameters, whereas $t_p$ is fixed at MJD 57189. The best fit values of these 
parameters with the corresponding reduced-$\chi^2$ and degrees of freedom ($\chi^2_r$/dof) for different light curves 
are summarized in Table \ref{tab:temp-par}. The values of the parameters $\tau_r$ and $\tau_d$ for the \emph{Fermi}-LAT 
light curve suggest that the $\gamma$-ray flare can be described by an exponential rise and decay with almost similar 
timescales of less than one day. But these fit parameters do not reproduce the two flux points measured in the high state 
before the peak of the flare. This indicates that the effective rise and decay times associated with the HE $\gamma$-ray 
flare peak are not similar. The 6 hours binning of the HE $\gamma$-ray flare peak suggests an exponential rise and decay 
times of $\sim$ 3 hours and $\sim$ 8 hours respectively [31]. The two flux points during the high activity state before 
the peak also have different rise and decay times [31]. Whereas, near simultaneous optical/IR emissions during the 
flaring episode are characterized by a relatively slow rise and very slow decay. 
The degree of symmetry in different light curves is determined by the time asymmetry parameter ($\zeta$) defined as [56]
\begin{equation}
		\zeta = \frac{\tau_d - \tau_r}{\tau_d + \tau_r}
\end{equation}
and the error in asymmetry parameter is given by 
\begin{equation}
	\Delta \zeta = \frac{2}{(\tau_d + \tau_r)^2} \sqrt{(\tau_d \Delta \tau_r)^2 + (\tau_r \Delta \tau_d)^2}
\end{equation}
where $\Delta \tau_r$ and $\Delta \tau_d$ are the uncertainties associated with $\tau_r$ and $\tau_d$ respectively.
The value of $\zeta$ lies between -1 and +1 which corresponds to right and left asymmetry of the flare respectively, 
whereas $\zeta = 0$ implies a time symmetric flare. We have estimated $\zeta$ for the HE $\gamma$-ray and optical 
flares using the values of $\tau_r$ and $\tau_d$ given in Table \ref{tab:temp-par}. For the HE $\gamma$-ray, 
the value of $\zeta$ is found to be -0.08$\pm$0.19, implying nearly time symmetric flare. However, symmetry in 
the HE $\gamma$-ray daily light curve is not statistically significant due to large uncertainty. 
This is consistent with the above observation that the HE $\gamma$-ray flare takes longer time to attain the peak 
followed by a relatively fast decay. But, optical emissions in different bands have an average value of 
$\zeta =$ 0.60$\pm$0.11, indicating a marked left asymmetric flare. This implies that the daily variation during the 
flaring activity in the optical bands is more asymmetric than that in the HE $\gamma$-ray regime. In the standard scenario, 
symmetric flares are associated with the crossing time of radiation or particles through the emission region whereas 
marked asymmetric flares are related to the fast injection and slow radiative cooling of the accelerated particles 
in the emission zone [56]. 
Therefore, the  time asymmetry parameter (although statistically insignificant) estimated in this study 
would suggest that the HE $\gamma$-ray flare can be attributed to the fast escape of radiation produced by the radiative 
cooling due to EC immediately after the injection of accelerated particles through the emission region in the case of 
leptonic process. However, optical flares indicate faster acceleration of particles and longer synchrotron cooling in 
the emission region of the jet. A time-dependent single zone leptonic and lepto-hadronic models have been tested to 
reproduce the observed hour-scale $\gamma$-ray variability of the FSRQ 3C 279 [33]. Both models face challenges in reproducing 
fully the characteristics of the historical flaring episode on June 16, 2015 (MJD 57189). Accelerating the escape of particles 
from the smaller emission regions in the leptonic model does not show any significant improvement in the fitting [33].   
\par
The measured degree of optical linear polarization in the wavelength range 400 nm-700 nm (Figure \ref{fig:Fig1} (c)) 
during this period also increases from $\sim$ 18$\%$ to 30$\%$. However, it attains the peak value on June 17, 2015 (MJD 57190), 
one day after the HE $\gamma$-ray flare and coincides with the optical flare which lasts for two days. 
Pittori et al. (2018) have also reported the similar measurement of the degree of polarization in R band with a 
maximum of 30$\%$ at  MJD 57190.2 [34]. This again indicates a delay of about 1 day after the peak of the HE $\gamma$-ray flare.
Such behaviour of the optical polarization had also been previously observed from the FSRQ 3C 279 during its HE $\gamma$-ray 
outburst in 2009 [8]. This particular HE $\gamma$-ray flare with a doubling timescale of approximately one day was observed to 
coincide with a significant decrease in the degree of optical polarization from 30$\%$ down to less than 10$\%$.  
In the present study, the degree of optical polarization peaks one day after the HE $\gamma$-ray outburst and then starts to decrease 
to a lower value. The increase in the  degree of linear polarization with near simultaneous enhancement in the optical emission in 
different bands supports its non-thermal origin due to the synchrotron process and ordering of the jet magnetic field in the emission 
region. The corresponding polarization angle shown in Figure \ref{fig:Fig1} (d) also increases gradually from a minimum of 25$^\circ$ to a 
maximum of 45$^\circ$ over a period of six days. The maximum value of the polarization angle is observed three days after the peak in the 
degree of polarization. The rise and decay of the degree of linear polarization in R band is accompanied by a rotation of 
polarization angle of $\sim$ 30$^\circ$ in 10 days [34]. This indicates that the orientation of the polarization vector slowly moves 
from north to east direction in the sky. The behaviour of the optical linear polarization is discussed in detail in Section 4.5.

\begin{table}
\caption{Parameters of the temporal profile (Equation \ref{eqn:temporal}) fitted to the HE $\gamma$-ray and 
	 optical light curves in  Figure \ref{fig:Fig1} (a) and (b).}
\begin{center}
\begin{tabular}{lcccccc}
\\
\hline
Light curve   	&$F_0$	&$F$	&$\tau_r$	&$\tau_d$	&$\chi^2_r/dof$\\
		&(erg cm$^{-2}$ s$^{-1}$)&(erg cm$^{-2}$ s$^{-1}$) &(days) &(days)	&\\
\hline
$\gamma$-ray	&(4.78$\pm$1.88)$\times10^{-9}$  &(1.69$\pm$0.23)$\times10^{-7}$  &(0.96$\pm$0.34) &(0.81$\pm$0.12)	&39/26\\
B     	 	&(5.88$\pm$0.73)$\times10^{-12}$ &(6.27$\pm$0.61)$\times10^{-12}$ &(2.53$\pm$0.98) &(17.34$\pm$6.86)	&219/13\\
V	 	&(6.76$\pm$0.75)$\times10^{-12}$ &(7.63$\pm$0.64)$\times10^{-12}$ &(2.56$\pm$0.88) &(18.04$\pm$6.43)	&188/15\\
R	 	&(7.46$\pm$0.83)$\times10^{-12}$ &(9.12$\pm$0.69)$\times10^{-12}$ &(2.48$\pm$0.85) &(19.09$\pm$6.38)	&229/15\\
J	 	&(1.27$\pm$0.61)$\times10^{-11}$ &(1.29$\pm$0.12)$\times10^{-11}$ &(1.88$\pm$0.55) &(10.26$\pm$1.98)	&103/12\\
\hline
\end{tabular}
\end{center}
\label{tab:temp-par}
\end{table}
\subsection{Variability}
To further quantify the temporal variability of the HE $\gamma$-ray and optical emissions from the FSRQ 3C 279 during the 
period June 1-30, 2015 including the flaring episode, two variability parameters are estimated. We have divided the entire 
period into three epochs: Pre-Flare (MJD 57174-57186), Flare (MJD 57187-57190) and Post-Flare (MJD 57191-57203). 
First, the fractional variability amplitude ($F_{var}$) is defined as [57]
\begin{equation}
	F_{var}=\sqrt{\frac{S^2 -E^2}{F^2}}
\end{equation}
and the error in $F_{var}$ is given by [57]
\begin{equation}
	\Delta F_{var}=\sqrt{\left(\sqrt{\frac{1}{2N}}\frac{E^2}{F^2F_{var}}\right)^2+\left(\sqrt{\frac{E^2}{N}}\frac{1}{F}\right)^2}
\end{equation}  
where $S^2$ is the variance, $E^2$ is the mean square measurement error, $F$ is the mean flux and $N$ is the number of flux points 
in a light curve. $F_{var}$ is a common measure of the amplitude of intrinsic variability present in the emission from a 
source after corrections for the effect of measurement uncertainties from an instrument. Second, the variability amplitude parameter 
($A_{mp}$) is calculated using the empirical formula [58]
\begin{equation}
	A_{mp}=100\times \frac{\sqrt{(F_{max}-F_{min})^2-2\sigma^2}}{F}~~\%
\end{equation}
and the uncertainty in $A_{mp}$ is given by [59]  
\begin{equation}
	\Delta A_{mp}=100\times \left(\frac{F_{max}-F_{min}}{FA_{mp}}\right)\sqrt{\left(\frac{\Delta F_{max}}{F}\right)^2 + 
	              \left(\frac{\Delta F_{min}}{F}\right)^2 + \left(\frac{\Delta F}{F_{max}-F_{min}}\right)^2 A_{mp}^4}~~\%
\end{equation}
where $F_{max}$ and $F_{min}$ are the maximum and minimum flux values with uncertainties $\Delta F_{max}$ and $\Delta F_{min}$ 
respectively in a light curve. $\Delta F$ is the error in the mean flux, and $\sigma$ is the average measurement error in the light curve.
$A_{mp}$ gives a measure of the percentage variation in the peak-to-peak flux points of the light curve. 
The values of two variability parameters $F_{var}$ and $A_{mp}$ estimated from the HE $\gamma$-ray and optical light curves for the 
whole period (MJD 57174-57203) and excluding the period of high activity (MJD 57187-57190) are reported in Table \ref{tab:var-par}. 
We observe that both variability parameters have much higher values when the flaring period is included in the HE $\gamma$-ray light curve 
than that for the low activity state of the source. However, a low activity state  HE $\gamma$-ray emission without a flaring episode also 
exhibits a variability which is consistent with the nature of the emission from the blazars. Also, we find that the HE $\gamma$-ray emission 
is more variable than the optical/IR emission in different bands. The values of $F_{var}$ and $A_{mp}$ obtained for the optical light 
curves are almost similar for the low activity state and including the flare period within statistical uncertainty. 
This is consistent with the observation that the highest flux points corresponding to the flaring episode in the optical light curves are 
not much higher than the flux measurements in the low activity state. Therefore, the parameters given in Table \ref{tab:var-par} quantify 
the variability in the HE $\gamma$-ray and optical emissions from the FSRQ 3C 279 as expected from the analysis of the light curves 
in Section 4.1.
\begin{table}
\caption{Parameters for quantifying the amplitude of temporal variability in the HE $\gamma$-ray and optical light curves.}
\begin{center}
\begin{tabular}{lccccc}
\\
\hline
Light curve		&\multicolumn{2}{c}{Entire Period}		&\multicolumn{2}{c}{Excluding Flare Period}\\
			&$F_{var}$	&$A_{mp}(\%)$			&$F_{var}$	&$A_{mp}(\%)$ \\
\hline
$\gamma$-ray		&2.01$\pm$0.25	&873$\pm$310			&0.68$\pm$0.11	&84$\pm$40\\
B     	        	&0.23$\pm$0.04  &69$\pm$4			&0.20$\pm$0.03	&72$\pm$4\\			
V	        	&0.24$\pm$0.04	&90$\pm$13			&0.23$\pm$0.04	&92$\pm$14\\
R	        	&0.21$\pm$0.03	&64$\pm$16			&0.21$\pm$0.04	&65$\pm$12\\
J	        	&0.24$\pm$0.04	&74$\pm$4			&0.19$\pm$0.03	&38$\pm$2\\	
\hline
\end{tabular}
\end{center}
\label{tab:var-par}
\end{table}

\begin{figure}
\begin{center}
\includegraphics*[height=0.49\textwidth,angle=-90]{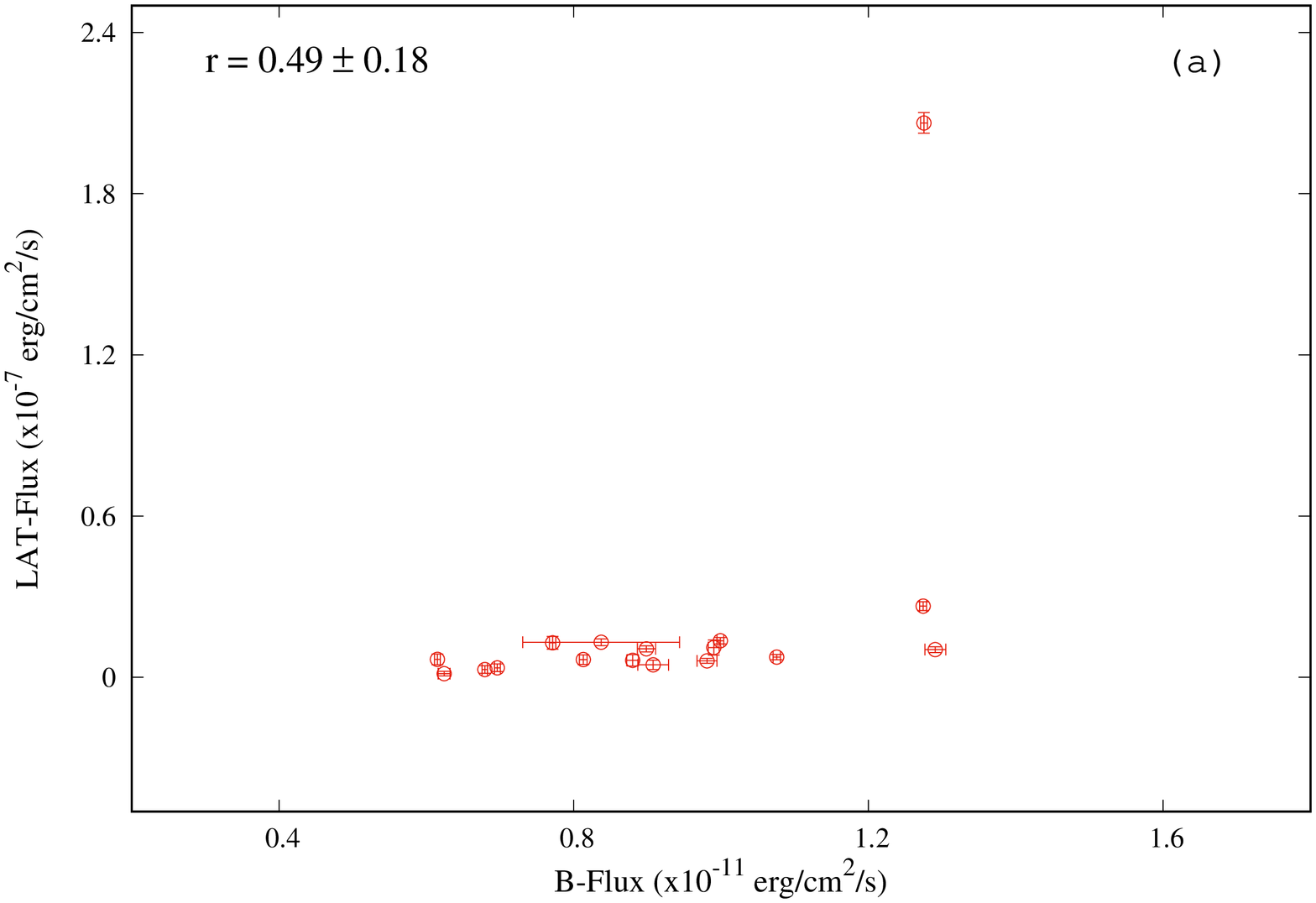}
\includegraphics*[height=0.49\textwidth,angle=-90]{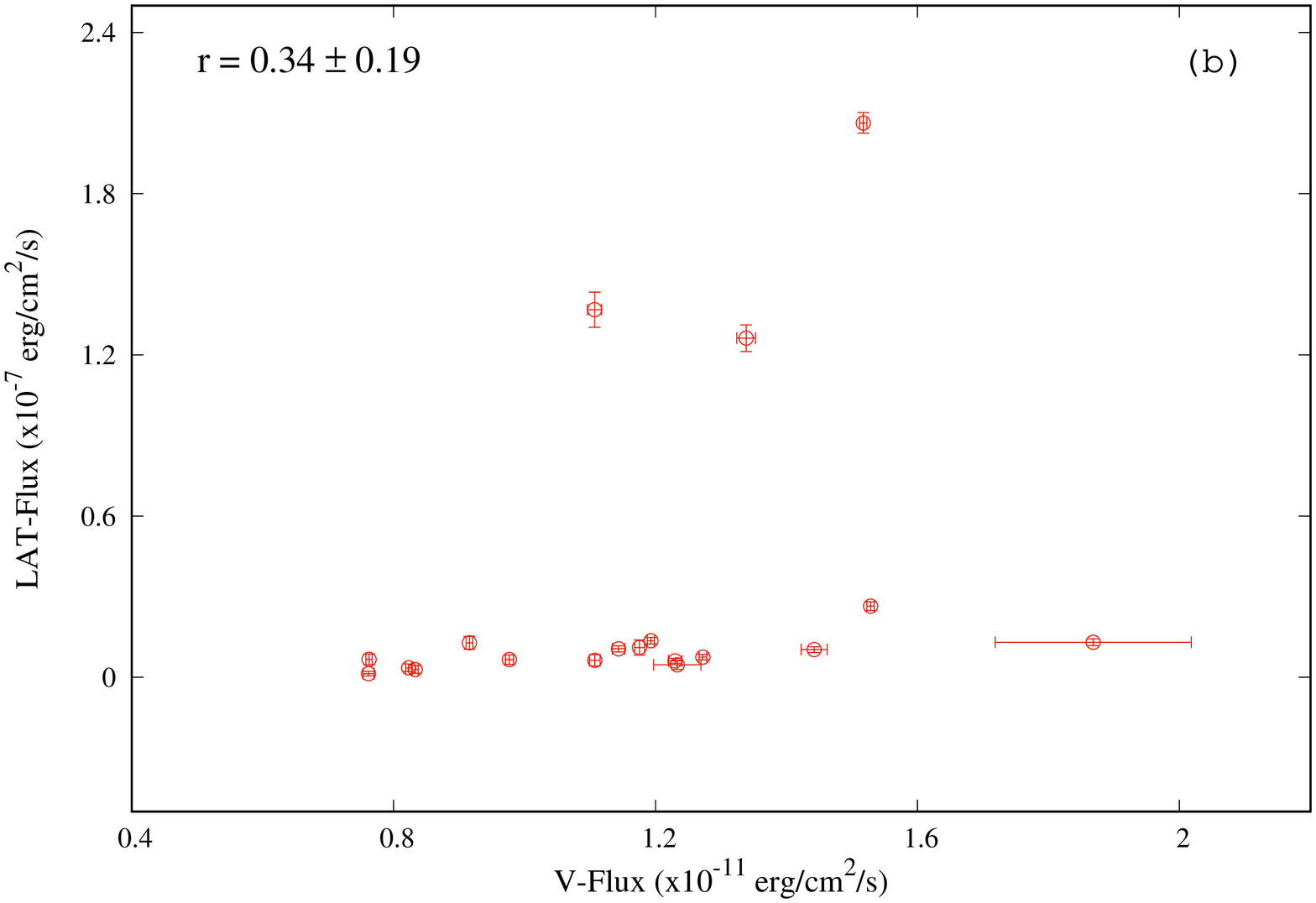}
\includegraphics*[height=0.49\textwidth,angle=-90]{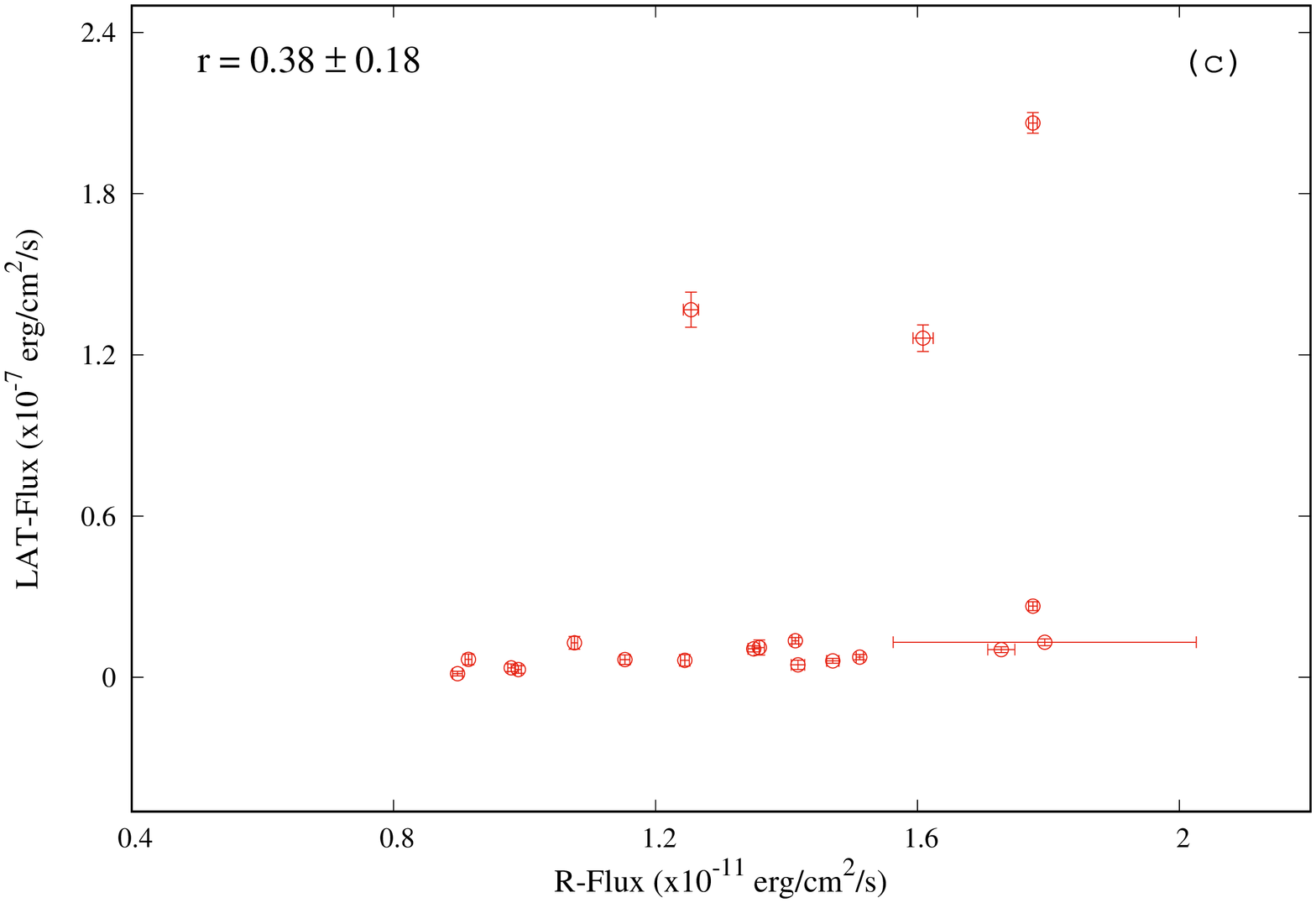}
\includegraphics*[height=0.49\textwidth,angle=-90]{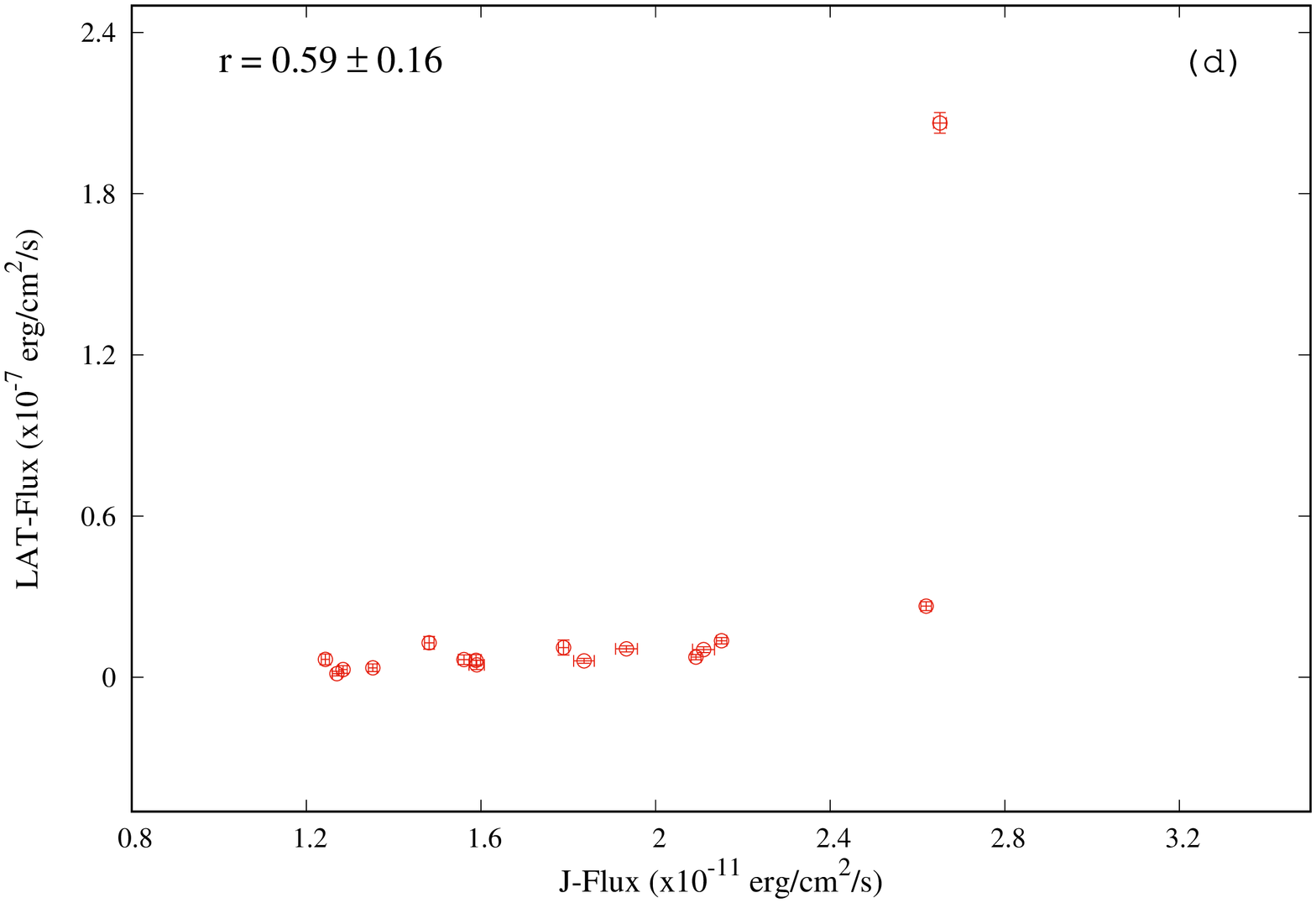}
\includegraphics*[height=0.49\textwidth,angle=-90]{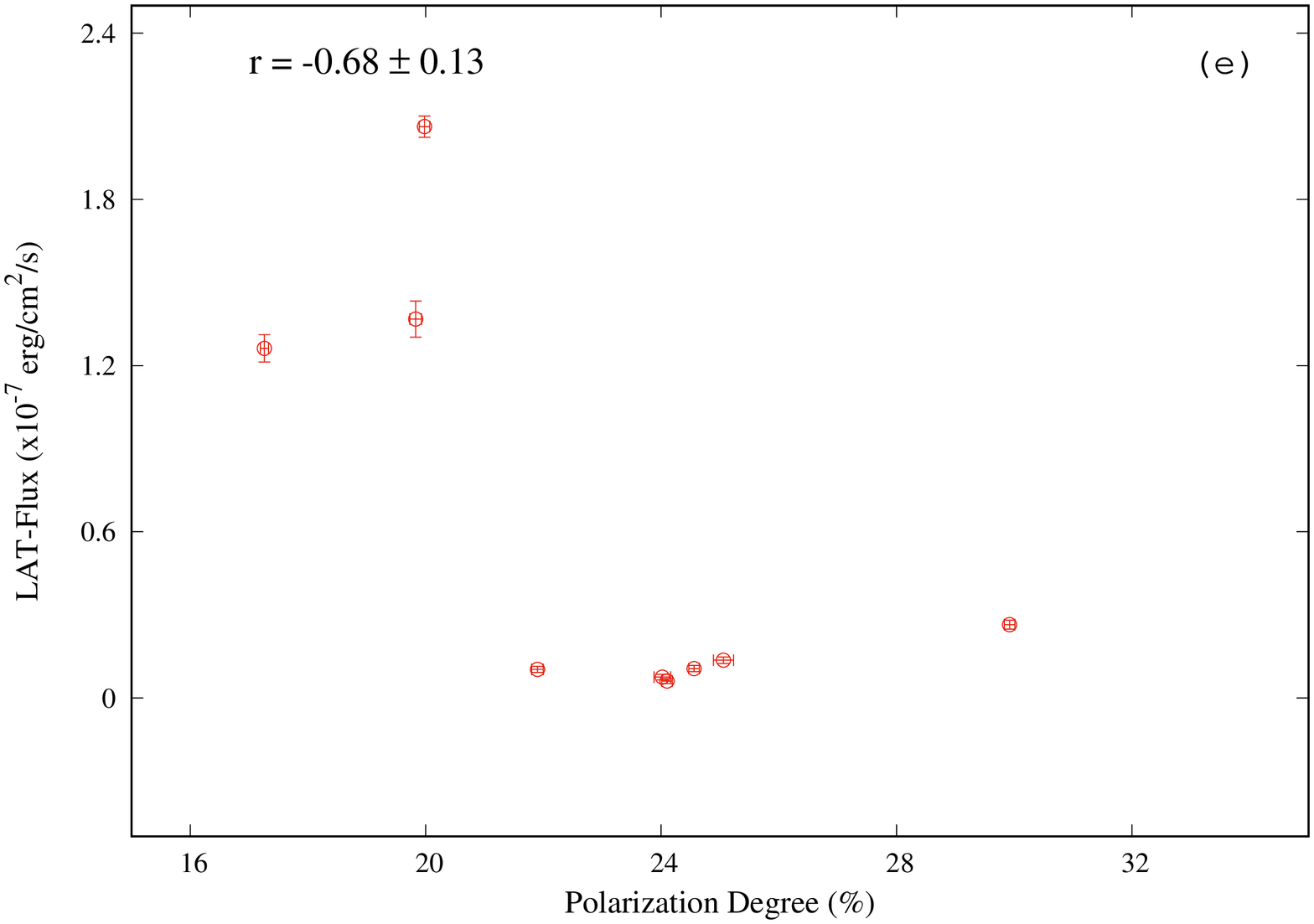}
\caption{Scatter plot for a linear correlation between near simultaneous HE $\gamma$-rays  and optical/IR observations.}
\label{fig:Fig2}
\end{center}
\end{figure}
\subsection{Correlation}
In order to investigate the physical connection between the HE $\gamma$-ray and optical/IR emissions, we have computed the 
Pearson correlation coefficient ($r$) and Spearman's rank correlation ($r_s$) using the \emph{Fermi}-LAT and SMARTS/SPOL 
measurements. A quantitative analysis based on Pearson coefficient provides a degree of linear correlation between emissions 
in two different energy bands. We have used near simultaneous flux points averaged over one day from the  \emph{Fermi}-LAT 
and SMARTS/SPOL observations during the period June 1-30, 2015 (MJD 57174-57203) to estimate the  Pearson correlation coefficient. 
The scatter plots for the HE $\gamma$-ray vs optical/IR emission in four bands (B, V, R, and J) with the corresponding values of 
the Pearson correlation coefficient ($r$) are shown in Figure \ref{fig:Fig2} (a-d). The values of the Pearson coefficient ($r$) 
indicate a moderate positive linear correlation between HE $\gamma$-ray and optical/IR emissions in different bands. This implies 
a plausible connection between the physical processes involved in the emission of multi-wavelength radiation from the source. 
The scatter plot between near simultaneous measurements of the degree of linear polarization and the HE $\gamma$-ray flux shown 
in Figure \ref{fig:Fig2} (e) indicates a negative  Pearson coefficient ($r=-0.68\pm0.13$). This suggests that the optical 
polarization in the wavelength range 400 nm-700 nm is anti-correlated with the HE $\gamma$-ray emission during the flaring episode. 
We have also estimated the Spearman's rank correlation ($r_s$) to further quantify the degree of correlation between the HE $\gamma$-ray 
and optical/IR emissions from the source. The computed values of $r_s$ using all near simultaneous flux points in the light curves and 
for the low activity state (excluding the highest flux points) are given in Table \ref{tab:spr-cor}. We observe that the HE $\gamma$-ray 
emission shows a moderate positive correlation with the optical emissions in B, V and R bands at a statistical significance level 
above 95$\%$. Whereas, a relatively strong positive correlation is found between the HE $\gamma$-ray and optical emission in J band 
with a statistical significance level of more than 98$\%$. Therefore, the Spearman's rank correlation analysis also confirms the 
connection between the HE $\gamma$-ray and optical emission processes in the jet of 3C 279.\\
The degree of linear polarization is anticorrelated with the HE $\gamma$-ray emission, but the statistical significance level is low. 
This behaviour is changed if the flaring episode is excluded and a positive correlation is observed. However, this positive correlation 
between the degree of polarization and HE $\gamma$-ray emission in the low activity state is not statistically acceptable due to very 
low statistical significance level and zero degree of freedom. 

\begin{table}
\caption{Spearman's rank correlation coefficient ($r_s$) between HE $\gamma$-ray and optical/IR light curves. 
	The correponding probabilities of the null-hypothesis (p-value) are computed for N-2 degrees of freedom (dof).}
\begin{center}
\begin{tabular}{lccccc}
\\
\hline
$\gamma$-ray    	&\multicolumn{2}{c}{Entire Period}		&\multicolumn{2}{c}{Excluding Flare Period}\\
vs			&$r_s$	&p-value				&$r_s$	&p-value \\
\hline
B     	        	&0.60   &0.05					&0.54	&0.05\\			
V		        &0.59	&0.02             			&0.51	&0.05\\
R		        &0.60	&0.02            			&0.50	&0.05\\
J	        	&0.72	&0.005           			&0.67	&0.02\\	
Polarization Degree     &-0.42  &0.50					&0.77	&0.20 (dof=0)\\
\hline
\end{tabular}
\end{center}
\label{tab:spr-cor}
\end{table}

\par
\begin{figure}
\begin{center}
\includegraphics*[height=0.49\textwidth,angle=-90]{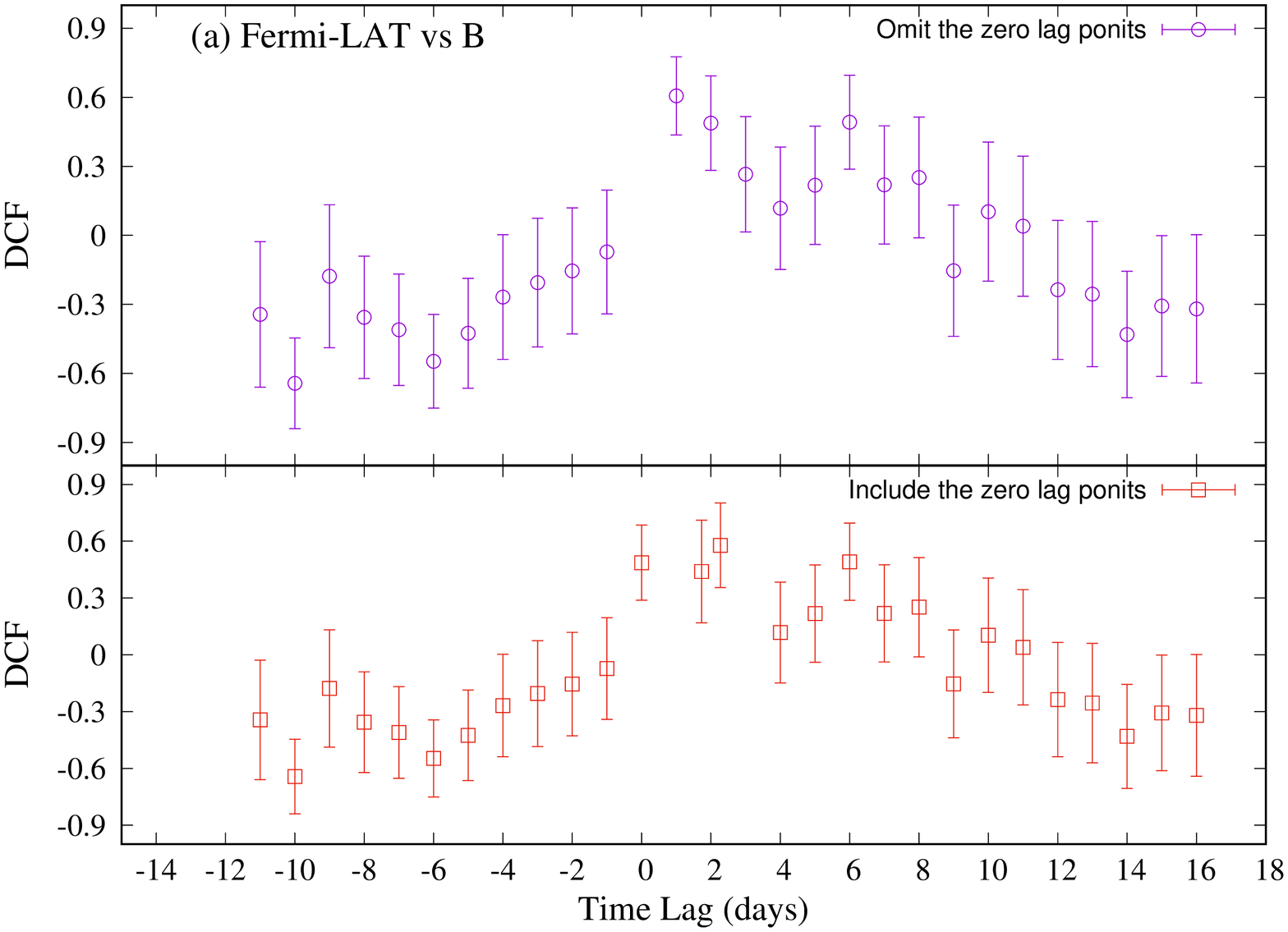}
\includegraphics*[height=0.49\textwidth,angle=-90]{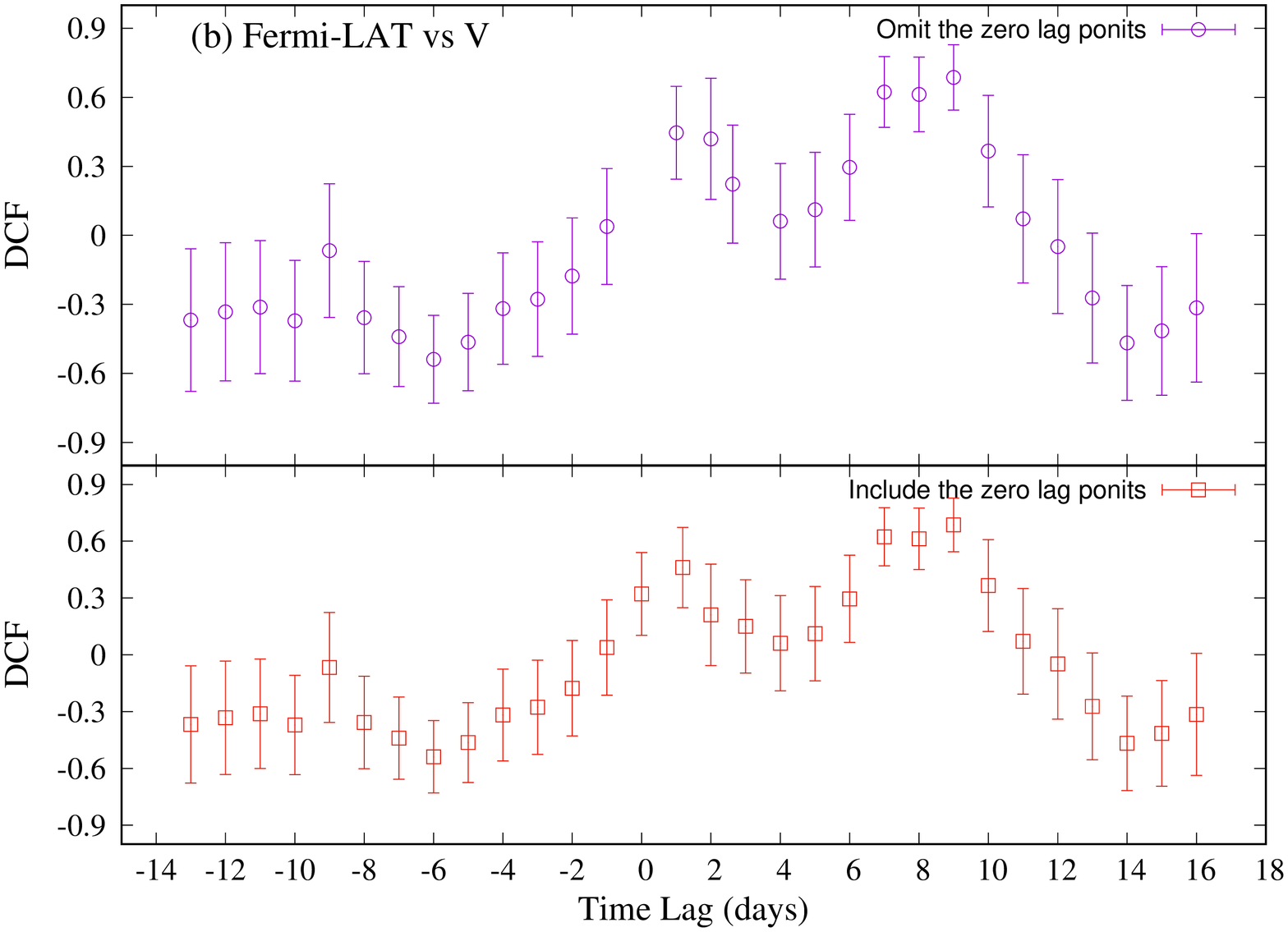}
\includegraphics*[height=0.49\textwidth,angle=-90]{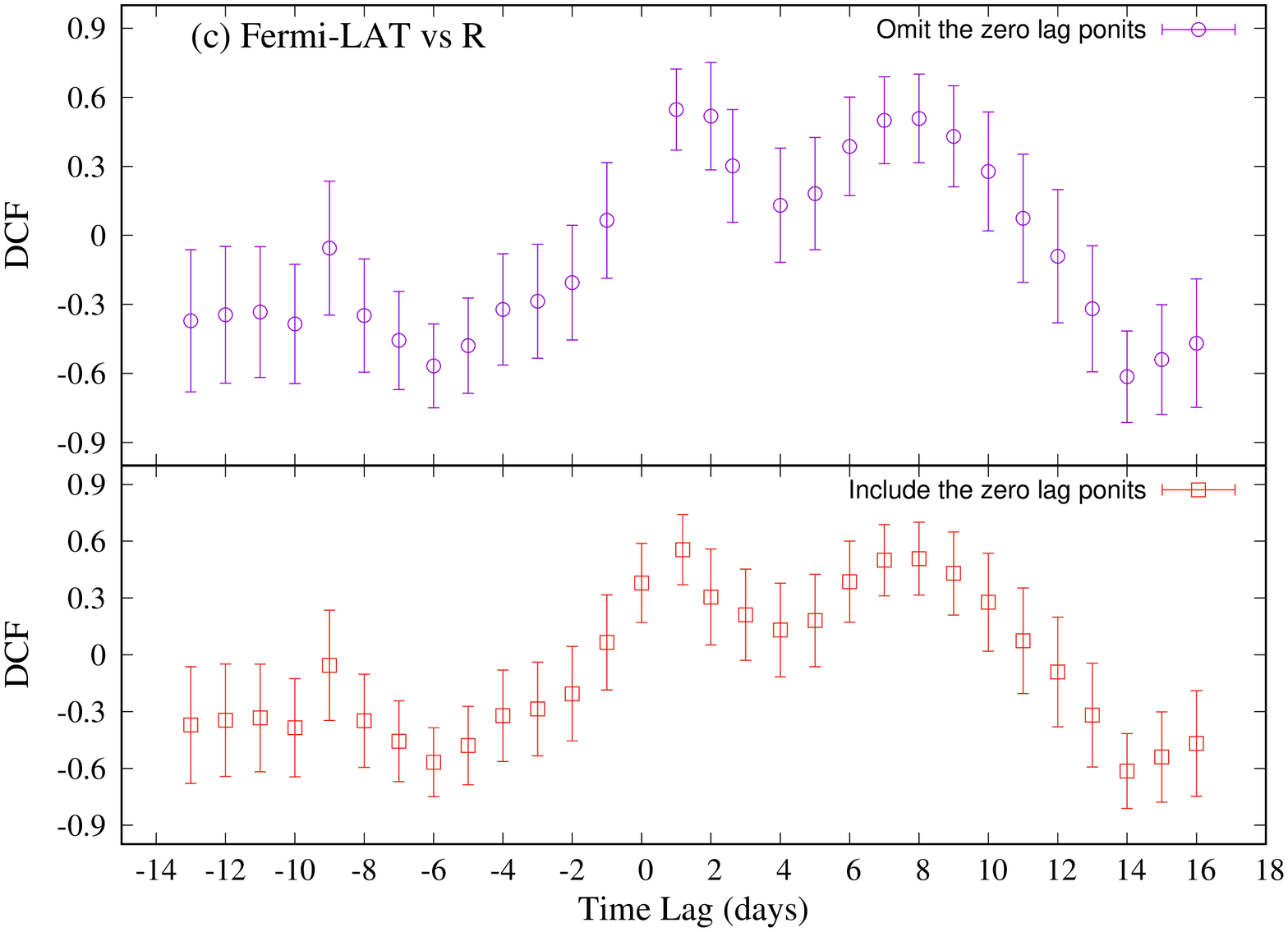}
\includegraphics*[height=0.49\textwidth,angle=-90]{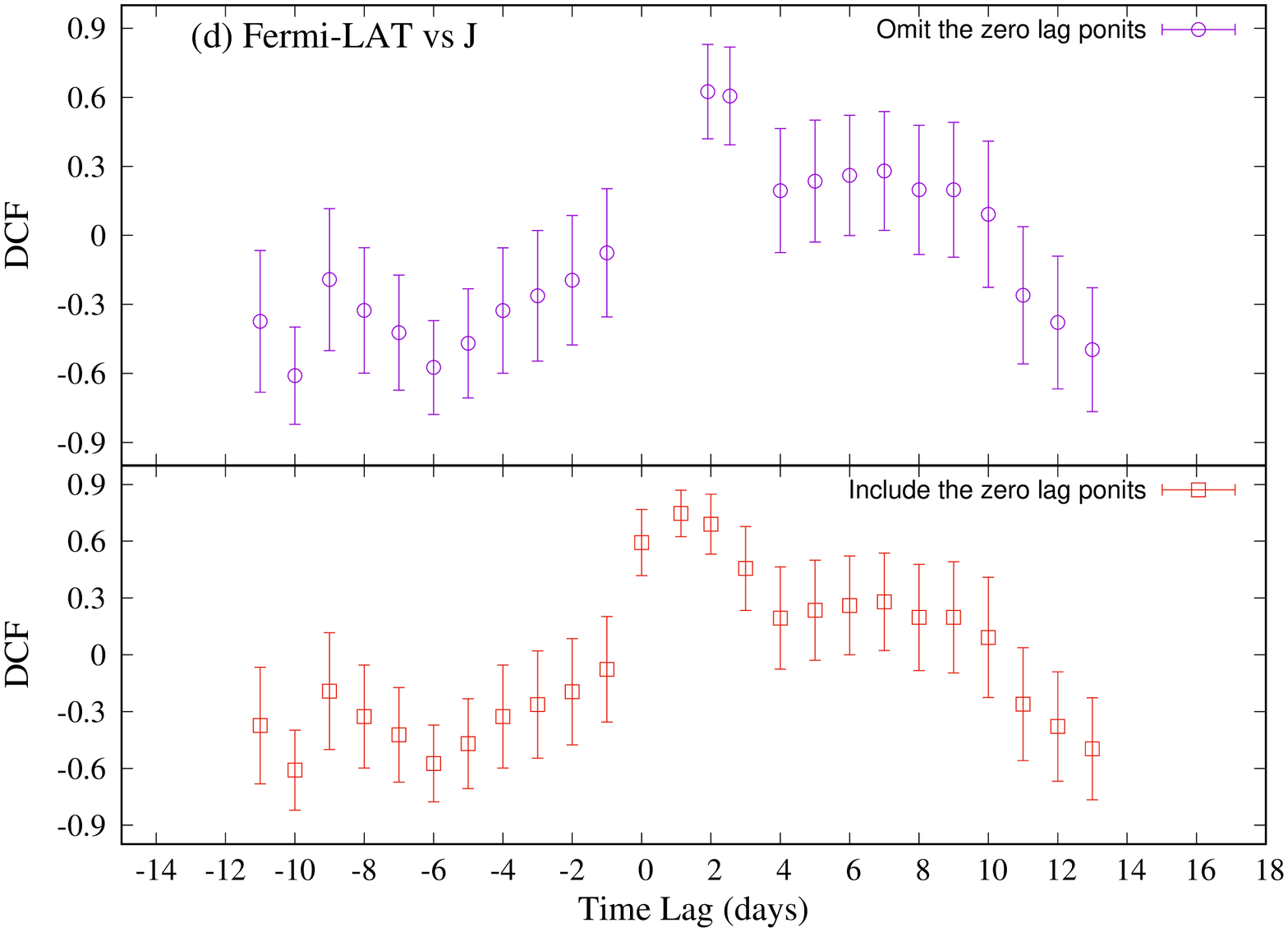}
\caption{ZDCF analysis curves showing the peak of correlation as a function of time-lag between the flares observed 
	in HE $\gamma$-ray and four optical/IR bands for two cases: Omit the zero lag points (open circles) and 
	Include the zero lag points (open squares).}
\label{fig:Fig3}
\end{center}
\end{figure}

In order to investigate the time-lags between the HE $\gamma$-ray and optical/IR flares in different bands, we have employed the 
z-transformed discrete correlation function (ZDCF) method fully described in [60]. The ZDCF-code is publicly and freely 
available for estimating the cross-correlation function of unevenly sampled 
light curves\footnote{http://www.weizmann.ac.il/particle/tal/research-activities/software}. 
We have considered a minimum of 11 points in each time bin while estimating the DCF coefficient for two different cases namely: 
omit the zero time lag points and include the zero lag points. The uncertainty in the estimated time-lag is computed using a maximum 
likelihood method [61]. The values of DCF and the corresponding time-lags in two different cases for the HE $\gamma$-ray and optical/IR 
emission in four bands are listed in Table \ref{tab:dcf1} (omit the zero lag points) and in Table \ref{tab:dcf2} (include the zero lag points). 
It is evident that the estimated DCF values for two different cases are almost similar within statistical uncertainties and 
the DCF values corresponding to the zero lag between the HE $\gamma$-ray and optical/IR light curves 
(for the case of include the zero lag points) are less statistically significant than the peaks at non-zero time-lags. 
The results of ZDCF analysis between $\gamma$-ray and optical/IR light curves in different bands for two different cases 
(omit and include zero lag points) are also shown in Figure \ref{fig:Fig3} (a-d). A prominent peak is observed in all the DCF curves 
at a time-lag of 1 day or longer. For V band, a time-lag of 9 days is due to the occurrence of a second peak on June 23, 2015 
(MJD 57196) in the light curve. As discussed in Section 4.1, the flux point associated with this peak has a relatively large error bar 
and therefore, can be considered as bad data. Excluding this data from the DCF calculations, we find that the DCF curves peak at 
a time-lag of 2 days and 1.8 days for the case of omit the zero lag points and include the zero lag points respectively.
The peak values of DCF indicate a positive correlation between the HE $\gamma$-ray and optical/IR light curves with emission 
in optical/IR bands lagging behind the $\gamma$-ray emission. This suggests that the optical/IR radiation is produced upstream of 
the HE $\gamma$-rays in the jet during the flare.   
\begin{table}
\caption{Results from the ZDCF analysis of the HE $\gamma$-ray light curve from the \emph{Fermi}-LAT vs optical/IR light 
	curves in four bands: Omit the zero lag points.}
\begin{center}
\begin{tabular}{lccccc}
\\
\hline
Optical/IR Band		&DCF 	&Time-Lag  	&Fiducial Interval	&Peak Likelihood\\
			&	&(days)		&(days)			&		\\			
\hline
B     			& $0.60_{-0.18}^{+0.15}$ &1.00	&(0.10, 5.70)  	&0.15\\			
V	        	& $0.68_{-0.15}^{+0.12}$ &9.00	&(6.95, 9.52)  	&0.14\\
R	        	& $0.54_{-0.21}^{+0.18}$ &1.00	&(0.08, 7.69)  	&0.07\\
J	        	& $0.62_{-0.22}^{+0.18}$ &1.90  &(0.61, 3.48)	&0.16\\	
\hline
\end{tabular}
\end{center}
\label{tab:dcf1}
\end{table}

\begin{table}
\caption{Results from the ZDCF analysis of the HE $\gamma$-ray light curve from the \emph{Fermi}-LAT vs optical/IR light 
	curves in four bands: Include the zero lag points.}
\begin{center}
\begin{tabular}{lccccc}
\\
\hline
Optical/IR Band		&DCF 	&Time-Lag  	&Fiducial Interval	&Peak Likelihood\\
			&	&(days)		&(days)			&		\\			
\hline
B     			& $0.57_{-0.24}^{+0.20}$ &2.27	&(0.43, 5.92)  	&0.11\\			
V	        	& $0.68_{-0.15}^{+0.12}$ &9.00	&(7.08, 9.53)  	&0.14\\
R	        	& $0.55_{-0.17}^{+0.20}$ &1.18	&(0.38, 8.03)  	&0.09\\
J	        	& $0.74_{-0.13}^{+0.10}$ &1.13  &(0.37, 2.18)	&0.18\\	
\hline
\end{tabular}
\end{center}
\label{tab:dcf2}
\end{table}

\subsection{HE $\gamma$-ray Spectrum}
\begin{figure}
\begin{center}
\includegraphics*[height=1.0\textwidth,angle=-90]{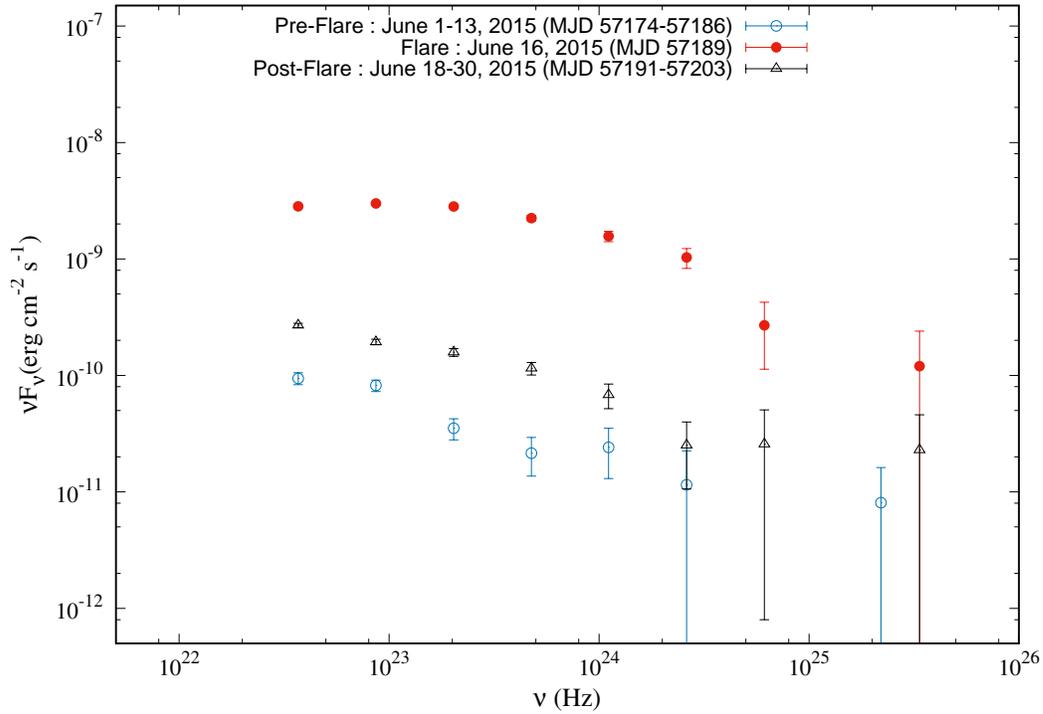}
\caption{HE $\gamma$-ray spectral energy distribution of the FSRQ 3C 279 in June 2015.}
\label{fig:Fig4}
\end{center}
\end{figure}

We have determined the time averaged HE $\gamma$-ray spectra of the FSRQ 3C 279 during three periods namely: Pre-Flare (MJD 57174-57186), 
Peak of the Flare (MJD 57189) and Post-Flare (MJD 57191-57203) in June 2015. The flux points measured by the \emph{Fermi}-LAT in different 
energy bands during the above three time periods are shown in Figure \ref{fig:Fig4}. The differential photon spectra in the energy 
range 0.1-500 GeV are fitted by a log-parabola (LP: Equation \ref{eqn:lp}) and a simple power law (PL: Equation \ref{eqn:lp} with $\beta = 0$). 
The results of the fits to the flux points in Figure \ref{fig:Fig4} for the three time periods are summarized in Table 
\ref{tab:gamma-spec}. We observe that the differential photon spectrum during the peak of the flare can be more favourably described 
by the LP model than the PL model due to relatively higher value of TS. However, during the pre and post-flare periods in the 
low activity state the spectra can be well described by a simple PL model as compared to the LP model since a change of 1 in TS values 
is not significant. The PL spectra during the pre and post-flare epochs indicate that the statistics may not be sufficient to detect the 
curvature ($\beta \sim$ 0) in the low activity state of the source. A comparison of the spectral flux points obtained during the pre 
and post-flare epochs suggests that the HE $\gamma$-ray emission level from the FSRQ 3C 279 after the decay of the outburst is relatively 
higher than that observed before the outburst. The spectral index $\alpha$ is observed to decrease with an increase in the emission level 
of the source. This indicates the presence of harder-when-brighter behaviour in the HE $\gamma$-ray emission from 3C 279 and the photon 
spectrum significantly changes from PL to LP when the source makes a transition from the low activity state to the flaring state. 
The peak energy ($E_p$) in the SED is related to the parameters of the LP model and can be expressed as [62]
\begin{equation}
	E_p = E_0 10^{(2 - \alpha)/2\beta}
\end{equation}	
From the values of the parameters which are reported in Table \ref{tab:gamma-spec}, we find that the peak position of the $\gamma$-ray 
component in the SED is at $\sim$ 0.27 GeV during the highest flaring episode. Whereas in the low activity states (pre and post-flare), 
$E_p$ is below 0.1 GeV because the power law spectral index $\alpha > $ 2. In the leptonic scenario for $\gamma$-ray emission from blazars, 
LP photon spectra hint that the relativistic electrons undergoing inverse Compton scattering in the emission region are also described by 
the LP model in a limited energy range [62]. However, the relations between the parameters of IC spectra and those of electrons are complex 
and depend on the dominance of scattering in Thomson or Klein-Nishina regime [62]. The curvature parameter ($\beta$) in the photon spectra 
mainly depends on the energy of seed photons, electron spectral curvature and energy of electrons participating in the IC scattering.
The HE $\gamma$-ray emission during the flaring activity of the source can be attributed to the upscattering of seed photons by a 
population of relativistic electrons described by LP energy distribution in the emission region. The LP energy distribution of the 
electrons can be produced by a statistical acceleration process in which the probability of acceleration decreases with the increase 
in electron energy [62]. Romoli et al. (2017) have shown that the time-averaged photon spectrum of 3C 279 during the 3 days of the 
flaring activity including the giant outburst on June 16, 2015 is described by a power law with a stretched exponential cut-off in 
the energy range 70 MeV to 300 GeV [63]. They suggest that the stretched cut-off in the $\gamma$-ray spectrum of 3C 279 during the 
flaring episode can be explained by a simple exponential cut-off in the spectrum of electrons involved in the EIC process or by 
the proton synchrotron process [63]. A modified or stretched exponential cut-off in the parent particle spectrum can be attributed to 
the interplay between radiative losses and acceleration of particles via diffusive shock acceleration or stochastic acceleration.

\begin{table}
\caption{Results from the HE $\gamma$-ray spectra of 3C 279 fitted using log-parabola and simple power-law during three 
	different epochs in the energy range 0.1-500 GeV.}
\begin{center}
\begin{tabular}{lccccccc}
\\
\hline
Period		&\multicolumn{4}{c}{Log-Parabola}		&\multicolumn{3}{c}{Power Law}\\
		&$\alpha$ &$\beta$ &$E_0$ (GeV) &TS 	&$\alpha$ &$E_0$ (GeV) &TS\\
\hline
Pre-Flare	&2.44$\pm$0.08 &0.05$\pm$0.06 &0.45 &507	&2.46$\pm$0.08	&1.4 &507\\
Flare-Peak     	&2.05$\pm$0.02 &0.11$\pm$0.01 &0.45 &27694	&2.13$\pm$0.02	&1.4 &27646\\			
Post-Flare	&2.30$\pm$0.03 &0.06$\pm$0.02 &0.45 &4354	&2.33$\pm$0.03	&1.4 &4353\\
\hline
\end{tabular}
\end{center}
\label{tab:gamma-spec}
\end{table}

\subsection{Optical Polarization}
\begin{figure}
\begin{center}
\includegraphics*[height=1.0\textwidth,angle=-90]{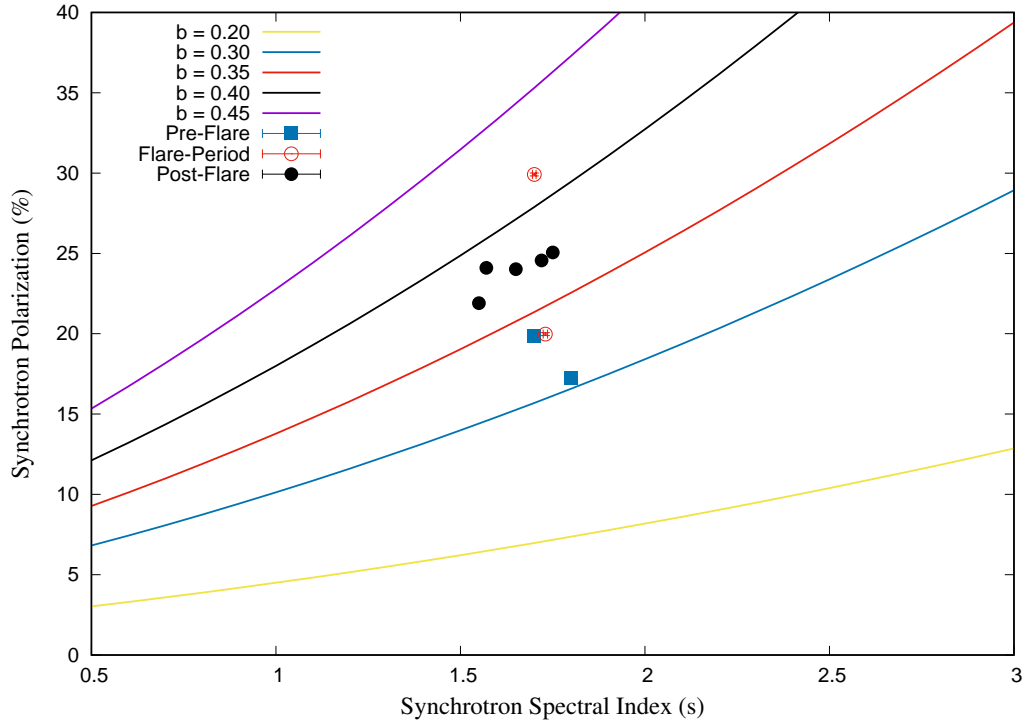}
\caption{Degree of linear synchrotron polarization as a function of synchrotron spectral index. The data points with blue and black 
	correspond to the daily measurements of the polarization during the Pre- and Post-Flare epochs respectively. The red data points 
	represent daily observations during the flare period. The highest degree of polarization ($\sim$ 30$\%$) has been measured one day 
	after the peak of HE $\gamma$-ray outburst.}
\label{fig:Fig5}
\end{center}
\end{figure}

The optical emission from the FSRQ class of blazars is observed to be highly polarized. Two important observables associated with 
the polarized emission from blazars namely degree of linear polarization ($P$) and angle of polarization ($\phi$) are found to 
exhibit variability just like broad-band radiation from these sources [64]. These observed features are in agreement with the 
synchrotron origin of the optical emission and indicate the presence of a partially ordered magnetic field in the jet. 
The observed variability in the synchrotron polarization can be attributed to the evolution of ordered magnetic field in 
the emission zone. In the simplest scenario, the magnetic field ($B$) in the synchrotron emitting region  can be described by 
two components: ordered ($B_o$) and tangled or chaotic ($B_c$). The degree of linear synchrotron polarization due to the 
resultant magnetic field ($B$) is given by [65,66]
\begin{equation}
		P_{sync} = \frac{(s+1)(s+2)(s+3)}{8(s+5/3)} b^2
\end{equation}	
where $b = \frac{B_o}{B_c}$ and $s$ is the synchrotron spectral index. The synchrotron spectrum which is observed in the 
optical/IR bands can be approximated by a simple power law of the form $F_\nu \propto \nu^{-s}$. We have used optical/IR 
flux measurements from SMARTS and SPOL observations of the FSRQ 3C 279 in four bands (B,V,R, and J) to derive the value 
of the synchrotron spectral index ($s$) in the flaring and low activity (excluding the flare) states. 
The time averaged values of $s$ for the Pre-Flare, Flare and Post-Flare epochs are 1.67$\pm$0.03, 1.71$\pm$0.02 and 
1.67$\pm$0.04 respectively. This implies that the optical spectrum softens during the flaring activity of the FSRQ 3C 279 and 
it remains the same in the low activity states before and after the flare. A comparison of the measured ($P$) and 
theoretically expected ($P_{sync}$) degree of linear polarization as a function of $s$ for daily observations 
corresponding to different values of $b$ is shown in Figure \ref{fig:Fig5}. We observe that the degree of polarization 
during the peak of HE $\gamma$-ray flare ($s \sim$ 1.73) is similar to that measured in the low activity state 
of the source before the flare (Pre-Flare epoch) and is significantly less than those during the Post-Flare epochs. 
During the Pre-Flare and peak of the HE $\gamma$-ray flare epochs, the measured degrees of linear polarization are broadly 
consistent with the theoretical synchrotron polarization for 0.30 $<$ b $<$ 0.35. Whereas, during the Post-Flare epoch, 
the observed degrees of polarization from the FSRQ 3C 279 are in agreement with the theoretical synchrotron polarization 
for 0.35 $<$ b $<$ 0.40. As mentioned in Section 4.1 and also reported in [34], the measured degree of linear polarization 
attains a peak of $\sim$ 30$\%$ nearly 1 day after the peak of HE $\gamma$-ray flare. This is found to be consistent 
with the theoretical synchrotron polarization for $b >$ 0.40. However, the intrinsic synchrotron polarization would be higher 
than the observed degree of polarization shown in Figure \ref{fig:Fig5} because thermal emission from the luminous accretion 
disk as well as stellar contribution from the host galaxy of 3C 279 can effectively depolarize the synchrotron polarization from 
the jet [65]. Therefore, taking into account the various depolarization effects during the comparison of the observed and 
theoretical synchrotron polarization, will give relatively larger values of $b$ (indicating the presence of a more 
ordered magnetic field in the optical emission region). Thus, we observe that the ratio of the strength of ordered magnetic 
field to the tangled magnetic field in the optical emission region decreases during the HE $\gamma$-ray outburst. This can 
be attributed to the evolution of the jet with distance because at large distances from the base of the jet, the magnetic field 
becomes tangled due to the motion of turbulent plasma [67]. This tangled magnetic field may be compressed 
by the shock waves to produce ordered magnetic field which results in the observed synchrotron polarization. 
The exact degree of ordering of the magnetic field can be determined from the combined optical, X-ray and $\gamma$-ray polarimetric 
measurements of the blazars [68]. Also, the gradual increase in the polarization angle (Section 4.1) indicates a non-axisymmetric 
magnetic field distribution in the jet or propagation of emission region along a curved trajectory [69]. The swing in polarization 
angle can also be caused by the freely propagating shocks in the jet. The FSRQ 3C 279 has been observed to exhibit different polarization 
variability characteristics during the low and flaring optical states governed by two different physical processes [39,70]. 
In the flaring state, the variability in polarization is attributed to a deterministic process whereas a stochastic process governs 
the variability in low activity state of the source.

\section{Active Region Parameters from observables}
One of the most interesting characteristics of the multi-wavelength observed emission from blazars is the temporal variability at 
timescales of minutes to several days. The most rapid variability with the shortest timescale allows to constrain some observables 
of the relativistic jet of a source. A well known relation for the size of the emission region ($R$) from where the fastest varying 
fluxes originate is given by
\begin{equation}
	R \le \frac{c~t_v~\delta}{1 + z}
\end{equation}	
where $t_v$ is the minimum observed variability timescale in the light curve, $\delta$ is the Doppler beaming factor and $c$ is the speed 
of light. From the daily light curves used in the present study, we find that the HE $\gamma$-ray emission from the FSRQ 3C 279 shows a
rapid variability with $t_v~ \sim$ 18 hours. This implies that the size of $\gamma$-ray emission zone can be constrained as 
\begin{equation}
	\frac{R}{\delta} \le 1.2\times10^{15}~\rm cm
\end{equation}
However, a detailed temporal analysis of the historical $\gamma$-ray flare on June 16, 2015 suggests a flux doubling time of less than 
5 minutes [32]. This minute scale variability further constrains the size of  $\gamma$-ray emission region as 
\begin{equation}
	\frac{R}{\delta} \le 5.8\times10^{14}~\rm cm
\end{equation}
which is an indicative of a very compact HE $\gamma$-ray emitting region in the jet of 3C 279 during the flare. For a relativistic 
jet with a viewing angle ($\theta$), the distance ($d$) of HE $\gamma$-ray flare emission zone from SMBH is given by 
\begin{equation}
	d = \frac{R}{\tan \theta}
\end{equation}
Radio observations of 3C 279 suggest $\theta~ \approx~2^\circ$ [25,26] and therefore for a minute scale HE $\gamma$-ray variability we get 
\begin{equation}
	d \le \delta \times 1.72\times10^{16}~\rm cm 
\end{equation}
The gravitational radius of SMBH in the FSRQ 3C 279 is $\sim$ 1.15$\times$10$^{14}$ cm. This implies that for a typical value 
of $\delta \sim$ 20, the distance of HE $\gamma$-ray flare zone from the SMBH is about 300 times the gravitational 
radius of SMBH [32,33]. This is two orders of magnitude less than the distance of dissipation region derived from an earlier 
$\gamma$-ray flare of the FSRQ 3C 279 observed in 2009 [8]. The location of the emission region obtained in the present study is also 
consistent with the estimated value by Paliya (2015) on the basis of the measured energy of the highest-energy photon detected during 
the flare on June 16, 2015 and $\gamma \gamma$ opacity arguments [31]. These estimates suggest that the HE $\gamma$-ray emission region 
during this historical flaring episode is located beyond the broad line emission region (BLR) in the jet of 3C 279 [31]. 
Results from the near simultaneous polarization measurements suggest a non-axisymmetric structure of the dissipation 
zone moving along a curved trajectory in the jet. The faster variability in the HE $\gamma$-ray emission compared to the optical in 
different bands implies that $\gamma$-rays are emitted from a region more compact than the optical emission zone during the outburst. 
The apparent isotropic HE $\gamma$-ray luminosity during the flare is given by [6]
\begin{equation}
	L_\gamma = \frac{F_P \times 4\pi d^2_L}{(1 + z)^{(2 - \alpha)}}
\end{equation}
where $F_P$ is the peak energy flux during the outburst and $d_L$ is the luminosity distance of the source. For the FSRQ 3C 279 with 
$d_L =$ 3.06 Gpc in the $\Lambda$-CDM cosmology, we get $L_\gamma ~\sim$ 2.36$\times$10$^{50}$ erg~s$^{-1}$ which is about three orders 
of magnitude higher than the Eddington luminosity ($L_{Edd} =$ 1.10$\times$10$^{47}$ erg~s$^{-1}$) for the mass of SMBH at the centre 
of the host galaxy. These values are consistent with the one obtained in other studies of the same flare [31,32,33]. 
This implies that the bolometric luminosity of the source during the flare is dominated by the HE $\gamma$-ray emission. 
Therefore, the estimated value of $L_\gamma$ can be used to constrain the total injected power ($P_{inj}$) associated with the 
relativistic electrons in the emission region ($P_{inj} \sim L_\gamma/\delta^4$) assuming that the entire injected power is 
converted into radiation in the fast cooling regime [71]. The total jet power will be close to or higher than the Eddington luminosity 
of the source.

\section{Conclusions}
We have performed a detailed temporal and spectral analysis of the historical outburst from the FSRQ 3C 279 which was 
observed on June 16, 2015 using near simultaneous data from the \emph{Fermi}-LAT, SMARTS and SPOL observations. 
The source shows a historical outburst with the HE $\gamma$-ray flux peaking at more than 25 times the flux level in the 
low activity state of the source in a time interval of less than one day. The peak of HE $\gamma$-ray flare in the daily 
light curve is characterized by a doubling timescale of $\sim$ 18 hours and a similar rise and decay timescales of $\sim$ 1 day. 
However, the asymmetry parameter estimated from the rise and decay timescales of the flare-peak is not statistically significant.
The optical/IR emissions in different bands also exhibit simultaneous increase in the flux but the change in the flux level 
during the flare with respect to the low activity state is not as dominant as in the HE $\gamma$-rays. 
The optical/IR flares are observed to be asymmetric with a shorter rise time and a larger decay time. 
The variability analysis of the near simultaneous HE $\gamma$-ray and optical light curves suggests that the $\gamma$-ray emission 
during the flare is more variable than the emissions in different optical/IR bands. A moderate and positive correlation is 
obtained between the HE $\gamma$-ray and optical/IR flux points which indicates a plausible connection between the physical 
processes associated with the emissions in the high and low energy bands. Further correlation analysis based on ZDCF-method also 
suggests a positive correlation between the HE $\gamma$-ray and optical/IR emissions with optical/IR flare lagging behind the 
$\gamma$-ray outburst by about one day or longer. 
The HE $\gamma$-ray emission during the flare is well described by a log-parabolic distribution whereas the emission in 
the low activity state is consistent with a simple power law. A harder-when-brighter behaviour is also observed in the HE $\gamma$-ray 
emission from the FSRQ 3C 279. The synchrotron spectra in the optical/IR band are broadly described by a simple power law with 
softer-when-brighter behaviour. The optical polarization is found to be anti-correlated with HE $\gamma$-ray flare 
with a significant drop in the degree of linear polarization in the wavelength range 400 nm -700 nm during an increase in 
the $\gamma$-ray band. The polarization behaviour of the source can be understood as the synchrotron polarization produced by 
the relativistic electrons in a magnetic field with an ordered and a chaotic components permeated in the emission region. 
The change in the degree of polarization is attributed to the interplay between the two components of the magnetic field. 
The HE $\gamma$-ray emission zone is compact and can be located at a distance 300 times the gravitational radius of SMBH 
in the jet of 3C 279 and the apparent isotropic $\gamma$-ray luminosity is much higher than the Eddington luminosity of the source.
A detailed broad-band SED modelling during the low and high activity states of the source is very important for the energetics of the 
relativistic jet of the FSRQ 3C 279.      
 
\section*{Acknowledgment}
We are very thankful to the anonymous reviewers for their important suggestions and comments to improve the manuscript.
We acknowledge the use of public data obtained through \emph{Fermi} Science Support Center (FSSC) provided by NASA. 
This paper has made use of up-to-date SMARTS optical/near-infrared light curves that are available at 
www.astro.yale.edu/smarts/glast/home.php. Data from the Steward Observatory spectropolarimetric monitoring project were used. 
This program is supported by Fermi Guest Investigator grants NNX08AW56G, NNX09AU10G, NNX12AO93G, and NNX15AU81G.


\end{document}